%% file: subclustering.tex
\documentclass[preprint, 11pt]{imsart}
\input{mypreamble.tex}

\begin {document}

\begin{frontmatter}
\title{Sub-clustering in decomposable graphs and size-varying junction trees}
\runtitle{Sub-clustering in decomposable graphs}
\author{\fnms{Mohamad} \snm{Elmasri}\corref{} \thanksref{a}\ead[label=e1]{mohamad.elmasri@mail.mcgill.ca}}
\runauthor{M. Elmasri}
\address{Department of Mathematics and Statistics \\ McGill University }

\thankstext{a}{ME was supported by the Fonds de recherche du Qu\'ebec - Nature et technologies (FRQNT) doctoral scholarship.\newline \printead{e1}}

\affiliation{McGill University}

\maketitle

\begin{abstract}
 This paper proposes a novel representation of decomposable graphs based on semi-latent tree-dependent bipartite graphs. The novel representation has two main benefits. First, it enables a form of sub-clustering within maximal cliques of the graph, adding informational richness to the general use of decomposable graphs that could be harnessed in applications with behavioural type of data. Second, it allows for a new node-driven Markov chain Monte Carlo sampler of decomposable graphs that can easily parallelize and scale. The proposed sampler also benefits from the computational efficiency of junction-tree-based samplers of decomposable graphs.
\end{abstract}


\end{frontmatter}

\begin{center}
  \begin{minipage}{.85\linewidth}
    \setcounter{tocdepth}{2}
    \tableofcontents
  \end{minipage}
\end{center}

\section{Introduction}

Statistical models using decomposable graphs have appeared in various topics and applications \citep{spiegelhalter1993,cowell2006probabilistic, Giudici01121999, frydenberg1989decomposition}. A primary use of decomposable graphs is in the context of graphical models, as a functional prior over covariance matrices. Given a data \(X\) with a conditional distribution abiding to a graph \(\G\) as \(p(X \mid \beta, \G)\), its prior distribution takes the form \(p(\beta \mid \G )p(\G)\). The interest is in specifying a prior over the space of \(\G\). For some appealing characteristics and computational efficiency, \(\G\) is often assumed to be decomposable. This work is also focused on decomposable structures, where we propose an alternative characterization of decomposable graphs based on bipartite graphs. The motivation is a flexible structure that could yield more efficient samplers.

\cite{Green01032013} proposed an efficient multi-edge Markov chain Monte Carlo (MCMC) sampler based on the junction tree representation, that improved on earlier samplers. Rather than selecting randomly sets of nodes to (dis)connect in a decomposable graph, the junction-tree-based sampler selects cliques and separators at random. It either, disconnects randomly a set of nodes within the clique, or connects nodes in adjacent cliques. This way, the sampler only constructs and updates a junction tree after every update move.

Motivated by \cite{Green01032013}, this work characterizes decomposable graphs as bipartite interactions between nodes and some latent communities representing the maximal cliques of the graph. The latent communities, having a tree-like structure, are only observable in the clique form by attaining node's memberships, analogous to the Indian Buffet Process \citep{Griffiths:2011:IBP:1953048.2021039}. In a sense, decomposable graphs are seen as a projective family from tree-dependent bipartite graphs, where the latter's
representation of decomposability is closer to the junction tree form. The first evident benefit of such characterization is that it allows for a notion of sub-clustering in maximal cliques. This could be leveraged in modelling behaviour type of data and in the MCMC sampler.

The bipartite representation allows for a node-driven MCMC sampler that enables easy parallel updates over different maximal cliques of the graph. The sampler is constructed for general use of decomposable graphs, but can easily be adapted for parameter-updating methods as in graphical models. As a result of the similarity between the bipartite representation and junction trees, the proposed sampler inherits some computational efficiency of the junction-tree-based sampler of \cite{Green01032013}.

This work is organized as follows: Section \ref{sec:preliminaries} introduces graph notations with a brief background on decomposable graphs, and an introduction to tree-dependent bipartite graphs. Section \ref{sec:clique-subgraphs-as} discusses a notion of sub-clustering in the bipartite representation, defines its junction graph, and proposes possible (dis)connect moves. Section \ref{sec:permissible-moves} illustrates the junction graph updates associated with each graph update. Section \ref{sec:Markov-size-varying-steps} gives a junction-graph-based MCMC sampler.

\section{Preliminaries}\label{sec:preliminaries}

\subsection{Notation and terminology} \label{sec:notation-terminology}
Let \(\G = (\Theta,E) \) be a simple undirected graph with a set of nodes \(\Theta = \{\theta_i\}_{i\in \N}\) and edges \(E = \{\{\theta_i, \theta_j\}\}_{i,j \in \N}\). A pair of nodes $\{\theta_i,\theta_j\} \in \Theta$ are adjacent if $\{\{\theta_i, \theta_j\}\}\in E$, or simply if \((\theta_i, \theta_j)\in E\) as \(\G\) is undirected. Let \(\G(x)\) defined a subgraph of \(\G\), such that, when \(x \subseteq \Theta\), then only edges connected to nodes in \(x\) are included, and when \(x \subseteq E\), only nodes forming edges in \(x\) are included. Let \(\neig\) be the operator returning the set of neighbouring nodes, such that, \(\neig(x, \G)\) are the neighbouring nodes of \(x\) in \(\G\) excluding those in \(x\), \(\neig(\G(x), \G)\) is an equivalent notation. Let \(\deg(x,\G)\) be the degree of node \(x\) in \(\G\). A subset $x \in \Theta$ is said to be complete if every two distinct nodes in $x$ are adjacent, and is commonly called a clique of \(\G\). Subgraphs of cliques are also cliques, thus, one can define a maximal clique to be a subgraph that cannot be extended by including any adjacent node while remaining complete. Finally, let \(\ver(x)\) be the set of nods associated with graph \(x\). 

\subsection{Decomposable graphs}\label{sec:decomposable-graphs}
The graph $\G$ is decomposable if, and only if, its maximal clique set \(\C\) can be ordered as \(\C_{\pi} = (C_{\pi(1)}, C_{\pi(2)}, \dots, C_{\pi(c)})\), for some permutation $\pi:\{1,\dots, c\} \mapsto \{1, \dots, c\}$, such that 
\begin{equation} \label{eq:perfectorder}
\text{for each}\quad S_{\pi(j)} = C_{\pi(j)} \cap \bigcup_{i=1}^{j-1} C_{\pi(i)},\quad S_{\pi(j)} \subset C_{\pi(k)} \text{ for }  k < j.
\end{equation}

The sequence of maximal cliques in \eqref{eq:perfectorder} is referred to as a perfect ordering sequence (POS), and the subset relation is known as the running intersection property (RIP) of the sequence. The set \(\S = \{S_{1}, \dots, S_{c}\}\) is called the minimal separators of $\G$, where each component in \(\S\) decomposes \(\G\) into subgraphs. While each maximal clique appears once in \(\C_{\pi}\), separators in \(\S\) could repeat multiple times, thus the naming of minimal separators as in the unique set of separators.

A decomposable graph \(\G\) can have multiple unique POSs, nonetheless, the sets \(\C\) and \(\S\) are unique. Enumerating all POSs of a graph is directly related to enumerating the set of junction trees spanning the graph. A tree $T=(\C, \Ep)$ is called a junction tree of $\G$, if the nodes of $T$ are the maximal cliques of $\G$, and the edges in \(\Ep\) correspond to $\S$. \cite{thomas2009a} have given and exact expression for the number of unique junction trees of a given decomposable graph. Moreover, the maximal cardinality search algorithm of \cite{Tarjan:1984:SLA:1169.1179} retrieves a junction tree representation in time order of \(|\Theta| + |E|\), where \(|\cdot|\) denotes the cardinality of a set. The junction tree concept is more general, that is, for any collection \(\C\) of subsets of a finite set of nodes of \(\Theta\), not necessary the maximal cliques, a tree \(T=(\C,\Ep )\) is called a junction tree if any pairwise intersection \(C_1 \cap C_2\) of pairs \(C_1, C_2, \in \C\) is contained in every node in the unique path in \(T\) between \(C_1\) and \(C_2\).

The interpretability of decomposability as conditional independence is the main drive of the statistical use of decomposable graphs. In particular, if a random variable \(X=(X_i)_{i<n}\) has a conditional dependency abiding to a decomposable graph \(\G\), then its likelihood factorizes as
\begin{equation}\label{eq:factordist}
  p(X\mid \G) = \frac{\prod_{C\in\C}p(X_C)}{\prod_{S\in \S} p(X_S)}.
\end{equation}
\nocite{dawid1993}

\subsection{Tree-dependent bipartite graphs}\label{sec:tree-depend-bipart}

Using the broader notion of junction trees, \cite{Elmasri2017a} defined a decomposable bipartite graph that maps to the classical representation of decomposable graphs in Section \ref{sec:decomposable-graphs}.

\begin{mydef}[tree-dependent bipartite graph] \label{def:tree-bi-graphs}
  Let \(\treebi = (\{\Theta',\Theta\}, E_{Z})\) be a bipartite graph connecting elements from the disjoint sets \(\Theta'\) and \(\Theta\). \(\treebi\) is a tree-dependent bipartite (tree-bi) graph if there exists a \(\Theta'\)-junction tree \(T=(\Theta', \Ep)\) of \(\treebi\).  That is, for any pair \(\theta'_{1}, \theta'_{2} \in \Theta'\), \(\neig(\theta'_{1}, \treebi) \cap \neig(\theta'_{2}, \treebi) \subseteq \neig(\theta'_{k}, \treebi)\), for every \(\theta'_{k}\) in the unique path in \(T\) between \(\theta'_{1}\) and \(\theta'_{2}\). 
\end{mydef}

The classical form of a decomposable graph \(\G\) can be achieved as projection from tree-bi graphs, defined as follows.
\begin{equation} \label{eq:mapping-adj-biadj}
  A = (a_{ij})_{ij} = \big ( \min\{\tilde{\z}^{\intercal}_{.i}\tilde{\z}_{.j},1\}\big)_{ij}, 
\end{equation}
where \(A\) is the adjacency matrix of \(\G\), and \(\tilde{\z}_{.j}\) is the \(j\)th row of \(\treebi\). Moreover, a junction tree of \(\G\) is a subtree of \(T\) of Definition \ref{def:tree-bi-graphs}. 

\begin{remark} Definition \ref{def:tree-bi-graphs} uses the notation \(\theta'_k\) interchangeably; for a subset of nodes of \(\treebi\), for a subset of nodes of \(\G\) representing the maximal cliques, and for the nodes in \(T\). To avoid ambiguity, let the term "node(s)" refer to the graph nodes, and "clique-node(s)" to the nodes in \(T\), that is in \(\Theta'\). For simplicity, we will often use the term "clique \(\theta'_k\)" to refer to nodes of the maximal clique in \(\G\) represented by \(\theta'_{k}\) as \(\neig(\theta'_{k}, \treebi)\). 
\end{remark}

Essentially, tree-bi graphs decouple the notion of cliques and nodes, where maximal cliques are assumed to be latent clique-communities that are observable in the \(\G\) form by attaining node's memberships. Hence, the junction tree of \(\G\) is a deterministic function of \(T\). 

The benefit of representing classical decomposable graphs in the tree-bi graph form \(\treebi\) is their simplified Markov update steps. \cite{Green01032013} illustrated a series of proposition addressing the Markov update rules for a decomposable graph \(\G\), for single and multi-edge updates. Although, \cite{Green01032013} update rules are more comprehensive, most of them can be abbreviated in a simple expression using \(\treebi\). Let \(\Tr{i}{}\) be the subtree of \(T\) induced by the node \(\theta_i\) as
\begin{equation} \label{eq:induced-tree}
  \Tr{i}{} = T^{}\Big(\{\theta'_s \in \Theta': (\theta'_{s}, \theta_i) \in E_{Z}\} \Big ).
\end{equation}

Then, the \(n+1\) Markov update step for \(\tilde{z}^{(n+1)}_{ki}\) conditional on the current configuration of \(T\) and \(\treebi^{(n)}\), including that of \(\tilde{z}^{(n)}_{ki}\) is 
\begin{equation}\label{eq:Update-step-simple}
\P(\tilde z^{(n+1)}_{ki} = 1 \mid \treebi^{(n)}, T)= 
\begin{cases}
 f(\theta'_k, \theta_i) & \text{if }\theta'_k \in \Tbd[(n)]{i}{}\bigcup \Tnei[(n)]{i}{}, \\ 
\tilde z^{(n)}_{ki} & \text{ otherwise,}
\end{cases}
\end{equation}
for some measurable function \(f: \Rp^{2}\mapsto [0,1]\). \(\Tbd{i}{}\) denotes the boundary clique-nodes of \(\Tr{i}{}\), those of degree 1 (leaf nodes), and \(\Tnei{i}{}\) the neighbouring clique-nodes in \(T^{}\) to \(\Tr{i}{}\), as
{\small
\begin{equation}\label{eq:nei-bound-clique-basic-no-cond}
\begin{aligned}
& \Tbd{i}{} := \Big \{\theta'_s \in \Theta': (\theta'_{s},\theta_{i})\in E_{Z}, \deg  (\theta'_s,\Tr{i}{})=1 \Big\},\quad \Tnei{i}{} := \neig(\Tr{i}{}, T)
\end{aligned}
\end{equation}
}
The model in \eqref{eq:Update-step-simple} is clearly iterative, updating \(\treebi \mid T\) and iteratively \(T\mid \treebi\). The model of \cite{Green01032013} is iterative as well, though their tree updates are coupled with local updates on the decomposable graphs. Refer to \citet[Sec. 3 \& 4]{Elmasri2017a} for more details.

\section{Clique subgraphs as sub-clusters}\label{sec:clique-subgraphs-as}

Tree-dependent bipartite graphs of Definition \eqref{def:tree-bi-graphs} have their benefits. However, the clique-nodes in \(\Theta'\) do not exclusively represent the maximal cliques of the mapped decomposable graph \(\G\); instead, they represent the maximal cliques and their sub-graphs \citep[Sec. 3 \& 4.1]{Elmasri2017a}. This is a direct result of the surjective mapping relation from \(\treebi\) to \(\G\) in the \(\Theta'\) domain. A bijective mapping is possible though with extra conditions imposed on \(\Tbd{i}{}\) and \(\Tnei{i}{}\) in \eqref{eq:nei-bound-clique-basic-no-cond} that cripples the update steps in (\ref{eq:Update-step-simple}). On the other hand, (\ref{eq:Update-step-simple}) is restrictive when \(\theta'_{k}\) is not maximal, since the decomposability constrain is not needed. In fact, nodes within a maximal clique can form multi-edges without restriction, as long as all edges are contained in a maximal clique. It can be argued as well, that updating \(\treebi\) with a dual update scheme, for maximal and sub-maximal cliques, can in fact accelerate convergence. 

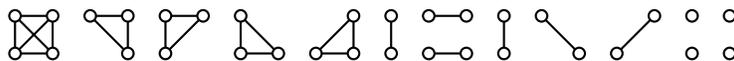
\begin{figure}[!ht]
  \centering
    \begin{tikzpicture}[scale=0.5, transform shape, thick]
      \node[circle,draw]  (a) at(0,0){};
      \node[circle,draw]  (b) at(0,1){};
      \node[circle,draw]  (c) at(1,0){};
      \node[circle,draw]  (d) at(1,1){};
      \draw (a)--(b) -- (c) -- (d) --(a) --(c);\draw (b)--(d);
      
      \node[circle,draw]  (b1) at(2,1){};
      \node[circle,draw]  (d1) at(3,1){};
      \node[circle,draw]  (c1) at(3,0){};
      \draw (b1)--(d1)--(c1) --(b1);

      \node[circle,draw]  (b2) at(4,1){};
      \node[circle,draw]  (d2) at(5,1){};
      \node[circle,draw]  (a2) at(4,0){};
      \draw (b2)--(d2)--(a2) --(b2);
      
      \node[circle,draw]  (b3) at(6,1){};
      \node[circle,draw]  (c3) at(7,0){};
      \node[circle,draw]  (a3) at(6,0){};
      \draw (b3)--(c3)--(a3) --(b3);

      \node[circle,draw]  (d4) at(9,1){};
      \node[circle,draw]  (c4) at(9,0){};
      \node[circle,draw]  (a4) at(8,0){};
      \draw (d4)--(c4)--(a4) --(d4);

      \node[circle,draw]  (a5) at(10,0){};
      \node[circle,draw]  (b5) at(10,1){};
      \draw (a5)--(b5);

      \node[circle,draw]  (d6) at(12,1){};
      \node[circle,draw]  (b6) at(11,1){};
      \draw (d6)--(b6);

      \node[circle,draw]  (a8) at(12,0){};
      \node[circle,draw]  (c8) at(11,0){};
      \draw (a8)--(c8);
      
      \node[circle,draw]  (d7) at(13,1){};
      \node[circle,draw]  (c7) at(13,0){};
      \draw (d7)--(c7);

      \node[circle,draw]  (b10) at(14,1){};
      \node[circle,draw]  (c10) at(15,0){};
      \draw (b10)--(c10);

      \node[circle,draw]  (d10) at(17,1){};
      \node[circle,draw]  (a10) at(16,0){};
      \draw (a10)--(d10);
      
      \node[circle,draw]  (a9) at(18,0){};
      \node[circle,draw]  (b9) at(18,1){};
      \node[circle,draw]  (c9) at(19,1){};
      \node[circle,draw]  (d9) at(19,0){};
    \end{tikzpicture}
\caption{A 4-node clique (left) and all its unique subgraphs for a total of 15.} 
\label{fig:cliques-and-subgraphs}
 \end{figure}

Implementing a dual update scheme, requires treatment of sub-maximal cliques. A node-labelled clique of size \(N\), has \(2^N-1\) unique subgraphs of smaller size cliques. Figure \ref{fig:cliques-and-subgraphs} illustrates an example of a 4-node clique with all its unique subgraphs forming smaller cliques, including single-node cliques. For simplicity, we will use the term "sub-clique(s)" to refer to sub-maximal clique(s), and "clique(s)" for maximal clique(s) unless otherwise specified.

 Accounting for all \(2^{N}-1\) unique subgraphs cliques requires a tremendous amount of bookkeeping that is deemed unnecessary. Instead, we adopt a representation analogous to that of the multi-graphs, where more than a single clique-node can represent the same unique sub-clique, prompting the importance of the latter. Therefore, at each Markov update step, it is only necessary to bookkeeping the set of clique-nodes (\(\C\)) representing maximal cliques. The relation between sub-cliques and their ascendant maximals is not exclusive, since sub-cliques within separators can be linked to multiple maximal cliques.

\begin{figure}[ht]
  \centering    
  \begin{minipage}[ht]{0.5\textwidth}
    \subfloat[][realization from \(\treebi\)]{\includegraphics[width=0.6\textwidth]{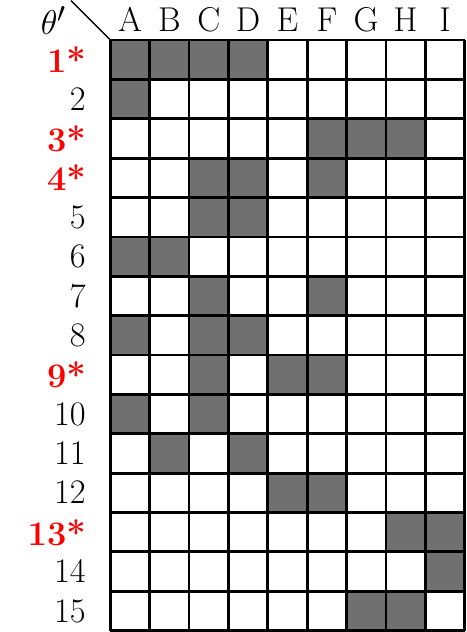}\label{fig:biadjacency-example-sub-clique-A}}
  \end{minipage}
  \hspace{-6em}
\begin{minipage}[ht]{0.5\textwidth}
\subfloat[junction graph \(T\)]{
  \begin{tikzpicture}[scale=0.4, transform shape, thick]
        \tikzstyle{every node}=[font=\large]
        \node[circle, draw, red] (abcd) at (0,0){ABCD};
        \node[circle, draw, red] (cdf) at (4,0){CDF};
        \node[circle, draw, red] (cef) at (8,0){CEF};
        \node[circle, draw, red] (fgh) at (12,0){FGH};
        \node[circle, draw, red] (hi) at (12,-3){HI};   
        \draw (2,0) -- (2,0) node[anchor=south]{CD};
        \draw (6,0) -- (6,0) node[anchor=south]{CF};
        \draw (10,0) -- (10,0) node[anchor=south]{F};
        \draw (12,-1.5) -- (12,-1.5) node[anchor=east]{H};
        \node[circle, draw, dashed] (ce) at (10,2){EF};
        \node[circle, draw, dashed] (cf) at (6,-2){CF};
        \node[circle, draw, dashed] (cd) at (2,2){CD};
        \node[circle, draw, dashed] (a) at (0,2){A};
        \node[circle, draw, dashed] (ab) at (-2,2){AB};
        \node[circle, draw, dashed] (d) at (2,-2){ACD};
        \node[circle, draw, dashed] (abc) at (-3,0){AC};
        \node[circle, draw, dashed] (bd) at (-2,-2){BD};
        \node[circle, draw, dashed] (gh) at (10,-2){GH};
        \node[circle, draw, dashed] (hi2) at (14,-1){HI};
        \draw [dashed] (cdf) - - (cd) -- (abcd) --(a);
        \draw [dashed] (ab) -- (abcd) -- (abc);
        \draw [dashed] (bd) -- (abcd) -- (d);
        \draw[dashed] (cdf) -- (cf) -- (cef)-- (ce);
        \draw (hi) -- (fgh) -- (cef) -- (cdf) -- (abcd);
        \draw[dashed] (hi2) -- (hi);
        \draw[dashed] (fgh) -- (gh);
       	\end{tikzpicture}\label{fig:biadjacency-example-sub-clique-C}}
      
      \subfloat[decomposable graph \(\G\)]{
        \begin{tikzpicture}[scale=0.7, transform shape, thick]
        \tikzstyle{every node}=[font=\large]
        \node[circle, draw] (a) at (0,0){A};
        \node[circle, draw] (b) at (0,2){B};
        \node[circle, draw] (c) at (2,2){C};
        \node[circle, draw] (d) at (2,0){D};

        \node[circle, draw] (e) at (4,2){E};
        \node[circle, draw] (f) at (4,0){F};

        \node[circle, draw] (g) at (6,2){G};
        \node[circle, draw] (h) at (8,0){H};
        \node[circle, draw] (i) at (8,2){I};
        
        \draw (a) -- (b) -- (c) --(a) --(d) --(c) --(f) -- (d) -- (b);
        \draw (c) -- (e) -- (f) -- (g) -- (h) --(f);
        \draw (h) -- (i);

        \draw (-1,2) --(-1,2) node[anchor=south] {\Large $\theta'_{\color{red} 1}$};        
        \draw (3,2) -- (3,2) node[anchor=south] {\Large $\theta'_{\color{red} 4}$};        
        \draw (3,0) --(3,0) node[anchor=north] {\Large $\theta'_{\color{red} 9}$};        
        \draw (6,0) --(6,0) node[anchor=north] {\Large $\theta'_{\color{red} 3}$};        
        \draw (8,1) --(8,1) node[anchor=west] {\Large $\theta'_{\color{red} 13}$};        
      \end{tikzpicture}\label{fig:biadjacency-example-sub-clique-B}}
  \end{minipage}
  \caption{An example of \(\treebi\) (left) with 5 maximal cliques, denoted by \(^{*}\) and in red, and 10 sub-cliques. The corresponding junction graph (top right) has all sub-cliques and their ascendants circulated and connected with dashed lines, with maximal cliques in red solid lines. Maximal cliques are represented in the corresponding decomposable graph (bottom right).}
    \label{fig:biadjacency-example-sub-clique}
\end{figure}
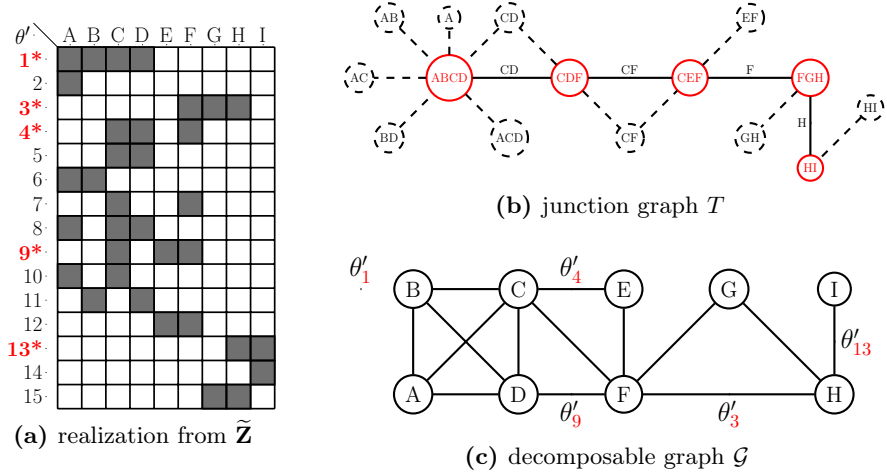

 Subfigure \ref{fig:biadjacency-example-sub-clique-A} shows a realization of \(\treebi\) with sub-maximal cliques, where only nodes participating in an edge are kept. The maximal cliques in \ref{fig:biadjacency-example-sub-clique-A} are denoted with \(^{*}\) and in red. The corresponding decomposable graph, shown in Subfigure \ref{fig:biadjacency-example-sub-clique-B}, consists of a 4-node, three 3-node, and a 2-node maximal cliques. Some sub-cliques are contained in multiple cliques, as shown with dashed lines in the junction graph of Subfigure \ref{fig:biadjacency-example-sub-clique-C}, where the sub-clique CD, also a separator, is contained in both ABCD and CDF.

The example in Figure \ref{fig:biadjacency-example-sub-clique} clearly demonstrates that sub-cliques in \(\treebi\) do not affect the decomposable graph directly, if disregarded, the graph is unchanged. Moreover, nodes can connect and disconnect to sub-cliques without risking decomposability, so long that all members of a sub-clique are also members of a single maximal clique.

Using different restrictions for maximal and sub-maximal cliques breaks the definition of \(T\) as a junction tree of the node set \(\Theta'\). In fact, \(T\) ceded to be a tree in Figure \ref{fig:biadjacency-example-sub-clique-C}, it is only a tree of a subset \(\C\) of \(\Theta'\) representing the maximal cliques. This slight difference in the meaning of \(T\), though subtle, has strong implications. Letting \(T\) be a latent \(\Theta'\)-junction tree, as in Definition \ref{def:tree-bi-graphs}, helped in decoupling the nodes \(\Theta\). In Figure \ref{fig:biadjacency-example-sub-clique}, \(T\) is a graph embedding the junction tree \(T_{\G}\) of \(\G\). This forces the \(\treebi\) updates to be local, similar to the proposal in \cite{Green01032013}, after every update in \(\treebi\) a corresponding update in \(T\) is needed.

To distinguish between the tree-dependent bipartite graph \(\treebi\) and the newly proposed representation, Definition \ref{def:clique-dependent-bipartite-graph} introduces a clique-dependent bipartite graph \(\Z\).

\begin{mydef}[clique-dependent bipartite graph] \label{def:clique-dependent-bipartite-graph}
  Let \(\Z = (\{\Theta',\Theta\}, E_{Z})\) be a bipartite graph connecting elements from the disjoint sets \(\Theta'\) and \(\Theta\), and let \(T = (\Theta', \Ep)\) be a graph connecting the elements of \(\Theta'\). Assume that \(\G\) is a decomposable graph formed by (\ref{eq:mapping-adj-biadj}), where \(\C\) is a subset of clique-nodes of \(\Theta'\) indexing the maximal cliques of \(\G\). \(\Z\) is a maximal clique-dependent (\(T_{\C}\)-dependent) bipartite graph if there exists a \(\C\)-junction tree \(T_{\C} = T(\C)\) of \(\Z\).  That is, for any pair \(\theta'_{1}, \theta'_{2} \in \C\), \(\neig(\theta'_{1}, \Z) \cap \neig(\theta'_{2}, \Z) \subseteq \neig(\theta'_{k}, \Z)\), for every \(\theta'_{k}\) in the unique path in \(T_{\C}\) between \(\theta'_{1}\) and \(\theta'_{2}\).
\end{mydef}

\begin{remark}
  The nodes of \(\Theta'\) represent the maximal cliques of \(\G\) and their sub-cliques. To avoid confusion, let \(\C\) represent the set of maximal cliques and \(\nC\) the set of sub-maximal cliques, such that, \(\Theta' = \C \cup \nC\). For clique ascendant relation, we use the subset notation, as \(x\subset y\) if \(x\) is a sub-clique of \(y\). For the set of parent cliques of a sub-clique we use the notation \(\pa(x):=\neig(x, T_{\C})\).
\end{remark}

The junction graph \(T\) of Definition \ref{def:clique-dependent-bipartite-graph} is not related to the concept of junction graph in \citet[Def. 2]{thomas2009a}.

Proposition \ref{prop:moved-on-Z} illustrates permissible single update moves in \(\Z\) ensuring it is a clique-dependent bipartite graph. The proposed moves are an adaptation of the efficient update conditions on decomposable graphs given by \cite{frydenberg1989decomposition,Giudici01121999}. That is, connecting any sets of nodes in a decomposable graph retains decomposability if, and only if, the sets of nodes are adjacent in some junction tree of the graphs. Disconnecting any sets of nodes retains decomposability if, and only if, the sets of nodes are contained in exactly one clique.

\begin{myprop}[Permissible moves in \(\Z\)] \label{prop:moved-on-Z} Following Definition \ref{def:clique-dependent-bipartite-graph}, let \(\Z\) be a clique-dependent bipartite graph with a \(\C\)-junction tree \(T_{\C} = T(C)\). For an arbitrary node \(\theta_{i} \in \Theta\), let \(\Tr{i}{}\) be the subtree of \(T\) induced by the node \(\theta_i\) as
  \begin{equation} \label{eq:induced-tree}
  \Tr{i}{} = T^{}\Big(\{\theta'_s \in \Theta': (\theta'_{s}, \theta_i) \in E_{Z}\} \Big ),\end{equation}
such that \(\Tr{i}{\C} = \Tr{i}{}(\C)\), the \(\theta_{i}\)-clique-subtree of \(T\). Moreover, let \(\Tbd{i}{}\) be the boundary clique-nodes of \(\Tr{i}{}\), those of degree 1 (leaf nodes) of some junction tree \(T_{\C}\), and \(\Tnei{i}{}\) the neighbouring clique-nodes in \(T^{}\) to \(\Tr{i}{}\), as

\begin{equation}\label{eq:nei-bound-sets}
  \begin{aligned}
    \Tbd{i}{\C} &:= \Big \{\theta'_s \in \C: \theta_{i} \in \theta'_{s}, \deg  (\theta'_s,\Tr{i}{\C})=1 \Big\},\\
    \Tbd{i}{\nC} &:= \Big \{\theta'_s \in \nC: \theta_{i} \in \theta'_{s} \Big\},\\
    \Tnei{i}{} &:= \neig(\Tr{i}{}, T),\quad  \Tnei{i}{\C} := \Tnei{i}{}\bigcap \C,\quad  \Tnei{i}{\nC} = \Tnei{i}{}\setminus \C
\end{aligned}
\end{equation}

Suppose \(\theta'_k\in \Tbd{i}{\C}\cup \Tbd{i}{\nC} \cup \Tnei{i}{} \), let \(\Z'\) be the graph formed by one of the following moves:
\begin{equation}
\begin{aligned}
    &\textbf{connect: } &z_{ki}=1,\quad & \text{if}\quad \theta'_{k} \in \Tnei{i}{}  \\
    &\textbf{disconnect: } &z_{ki}=0,\quad & \text{if}\quad \theta'_{k} \in \Tbd{i}{\C}\cup \Tbd{i}{\nC}.
  \end{aligned}
\end{equation}

Then, \(\Z'\) is also a clique-dependent bipartite graph.
\end{myprop}

\begin{remark} The notation \(\neig(\Tr{i}{}, T)\) in \eqref{eq:nei-bound-sets} includes cliques of disconnected components in \(T\) and sub-clique of neighbouring maximal clique that satisfy \(\{x\in \nC: \pa(x)\cap \Tr{i}{\C} \subset x\}\).
\end{remark}

The proposed representation of decomposable graphs as clique-dependent graphs is also a model based on junction trees. Moreover, it allows for sub-clustering formation within maximal cliques, that will be used to leverage some efficiency in the Markov chain sampler alongside the efficiency gained from the junction tree representation. Like the work of \cite{Green01032013}, the following section illustrates required updates to the junction tree after a node's perturbation.

\section{Junction graph updates in clique-dependent bipartite graphs}\label{sec:permissible-moves}

Following the settings of Definition \ref{def:clique-dependent-bipartite-graph} and Proposition \ref{prop:moved-on-Z}, the set of permissible moves in \(\Z\) is organized into three parts: the connect, the promotion to maximal, and the disconnect move.

\subsection{The connect move}\label{sec:connect-node}

From Proposition \ref{prop:moved-on-Z}, nodes connect to cliques or sub-cliques that are adjacent in \(T\) to their induced \(\theta_{i}\)-subtree \(\Tr{i}{}\), including disconnected components. While the update move is simple, the modification of the junction graph \(T\) is more complicated. Starting with \(T = (\Theta', \Ep)\), where \(\Theta'=\C\cup \nC\), connecting \(\theta_{i}\) to some \(\theta'_{s}\in \Tnei{i}{}\) in \(\Z\) would result in the junction graph \(T'= (\C'\cup \nC', \Ep')\), with \(T'(\C')\) being the junction tree, by the following modifications:
\begin{itemize}[leftmargin=*]
\item in case \(\theta'_{s} \in \nC\): 
  \begin{enumerate}[(a)]
  \item if \(\theta_{i}\in \pa(\theta'_{s})\): remove all edges \(\{(x, \theta'_{s}) \in \Ep: \theta_{i}\not\in x \}\).
  \item if \(\theta_{i}\not\in \pa(\theta'_{s})\): 
    \begin{enumerate}[i.]
    \item remove the clique-nodes \(\{x \in \ver(\Tr{i}{\C}):x\subset \theta'_{s}\cup\{\theta_{i}\}\} \) from \(\C\);
    \item replace the junction tree edge \(\{(x, \pa(\theta'_{s}))\in \Ep: x \in \ver(\Tr{i}{\C})\}\), if exists, with \((y, \theta'_{s})\) for \(y \in \Tr{i}{\C}\) such that \(\theta'_{s}\cap \Tr{i}{\C} \subset y\);
    \item remove all edges \(\{(x, \theta'_{s})\in \Ep\}\) except one and the edge in ii.;
    \item add edges \((x, \theta'_{s})\) to \(\Ep\) for  \(\{x \in \neig(\pa(\theta'_{s}), T): x\subset \theta'_s \cup \{\theta_i\}\}\).
    \end{enumerate}
  \end{enumerate}
  \item in case  \(\theta'_{s} \in \C\): follow (b)\{i., ii., iv.\} above. 
  \end{itemize}

  Note that the steps in (b) are performed sequentially, and at times the conditions can result in an empty set, for example, in (b).iii when a sub-clique has only one edge. The conditions above match to some extent the multi-edge connect moves in \citet[Sec 3.1]{Green01032013} when \(\theta'_{s}\) is maximal. They correspond to the single-edge connect move of \cite{Green01032013} when \(\theta'_{s}\) is a single-node clique, to the multi-edges connect move when \(\theta'_{s}\) is a multi-node clique.
  
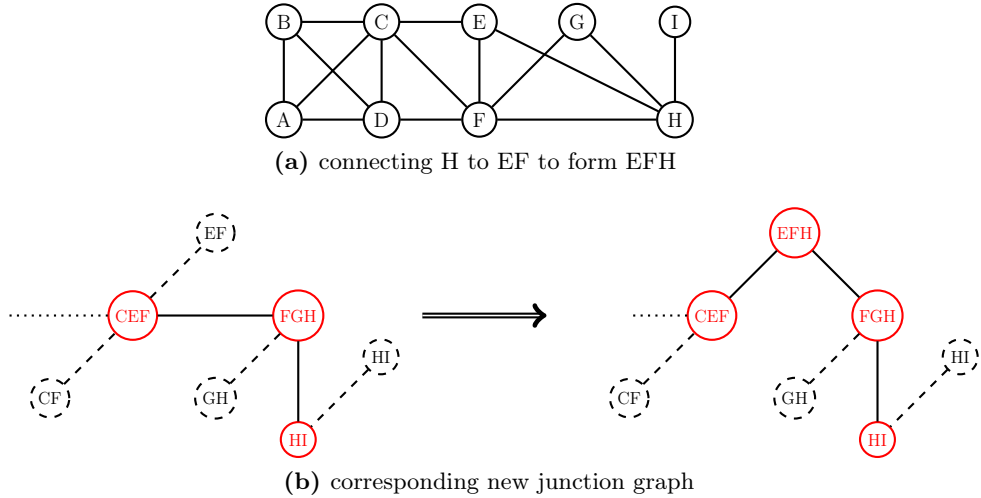
\begin{figure}[h!t]
  \centering    
  \subfloat[connecting H to EF to form EFH]{
    \begin{tikzpicture}[scale=0.65, transform shape, thick]
        \tikzstyle{every node}=[font=\large]
        \node[circle, draw] (a) at (0,0){A};
        \node[circle, draw] (b) at (0,2){B};
        \node[circle, draw] (c) at (2,2){C};
        \node[circle, draw] (d) at (2,0){D};
        
        \node[circle, draw] (e) at (4,2){E};
        \node[circle, draw] (f) at (4,0){F};

        \node[circle, draw] (g) at (6,2){G};
        \node[circle, draw] (h) at (8,0){H};
        \node[circle, draw] (i) at (8,2){I};
        
        \draw (a) -- (b) -- (c) --(a) --(d) --(c) --(f) -- (d) -- (b);
        \draw (c) -- (e) -- (f) --(g) --(h) -- (f);
        \draw (i) --(h) --(e);

        \node (i1) at (8,2){};
      \end{tikzpicture}}
    
    \subfloat[corresponding new junction graph]{
  \begin{tikzpicture}[scale=0.55, transform shape, thick]
        \tikzstyle{every node}=[font=\large]
        \node[circle, draw, red] (cef1) at (-6,0){CEF};
        \node[circle, draw, red] (fgh1) at (-2,0){FGH};
        \node[circle, draw, red] (hi1) at (-2,-3){HI};   

        \node[circle, draw, dashed] (gh1) at (-4,-2){GH};
        \node[circle, draw, dashed] (ce1) at (-4,2){EF};
        \node[circle, draw, dashed] (cf1) at (-8,-2){CF};
        \node[circle, draw, dashed] (hi21) at (-0,-1){HI};
        
        \draw [dashed] (cf1) -- (cef1)-- (ce1);
        \draw (hi1) -- (fgh1) -- (cef1);
        \draw[dotted] (cef1) -- (-9,0);
        \draw[dashed] (hi21) -- (hi1);
        \draw[dashed] (fgh1) -- (gh1);
        
        \draw[ -> , thick, double] (1,0) -- (4,0);
        \node[circle, draw, red] (cef) at (8,0){CEF};
        \node[circle, draw, red] (fgh) at (12,0){FGH};
        \node[circle, draw, red] (hi) at (12,-3){HI};   
        \node[circle, draw, dashed] (gh) at (10,-2){GH};

        \node[circle, draw, red] (ce) at (10,2){EFH};
        \node[circle, draw, dashed] (cf) at (6,-2){CF};
        \node[circle, draw, dashed] (hi2) at (14,-1){HI};
        \draw [dashed] (cf) -- (cef);
        \draw (hi)-- (fgh) -- (ce) --(cef);
        \draw[dotted] (cef) -- (6,0);
        \draw[dashed] (hi2) -- (hi);
        \draw[dashed] (fgh) --(gh);
      \end{tikzpicture}}
    \caption{An example of connecting a node to a sub-clique in an adjacent maximal clique. Node H connects to the sub-clique EF (left), from the example in Figure \ref{fig:biadjacency-example-sub-clique}, forming the new maximal clique EFH. The junction graph is achieved by following (b)\{i. to iv.\} of Section \ref{sec:connect-node}.}
    \label{fig:connecting-node-to-sub-clique}
\end{figure}

  Illustrative examples of the case of \(\theta'_{s}\in \nC\) and (a) can easily be conceived from the example in Figure \ref{fig:biadjacency-example-sub-clique}. The case \(\theta'_{s}\in \C\) is extensively illustrated in \citet[Fig. 3]{Green01032013}. Hence, Figure \ref{fig:connecting-node-to-sub-clique} illustrates the case of \(\theta'_{s}\in \nC\) and (b); connecting a node to an adjacent sub-clique in Figure \ref{fig:biadjacency-example-sub-clique}.

\subsection{Sub-cliques as multi-edges} \label{sec:sub-cliques-as-multi-edges}

In the classical representation of decomposable graphs, \citet{Green01032013} illustrated multi-edge (dis)connect moves that preserve decomposability. Such moves, can be mapped, more of less, to update moves on \(\Z\). Nonetheless, the interpretation of disconnect moves in \(\Z\) differs from those in \cite{Green01032013}. In their work, and some others, disconnecting two sets of nodes is identical to removing all edges connecting the two sets. In multi-graphs, nodes are able to form multi-edges in between; hence, it is possible to assume a probability model disconnecting a single, a fraction, or all the multi-edges between a set of nodes. The sub-clique representation in \(\Z\) mimics that of the multi-edge interpretation in multi-graphs, where disconnecting a set of nodes does not imply severing all multi-edges in a maximal clique. Nonetheless, multi-edge disconnect moves are needed to preserve the decomposability of \(\Z\).

\begin{figure}[ht!]
  \centering    
  \includegraphics[width=0.7\textwidth]{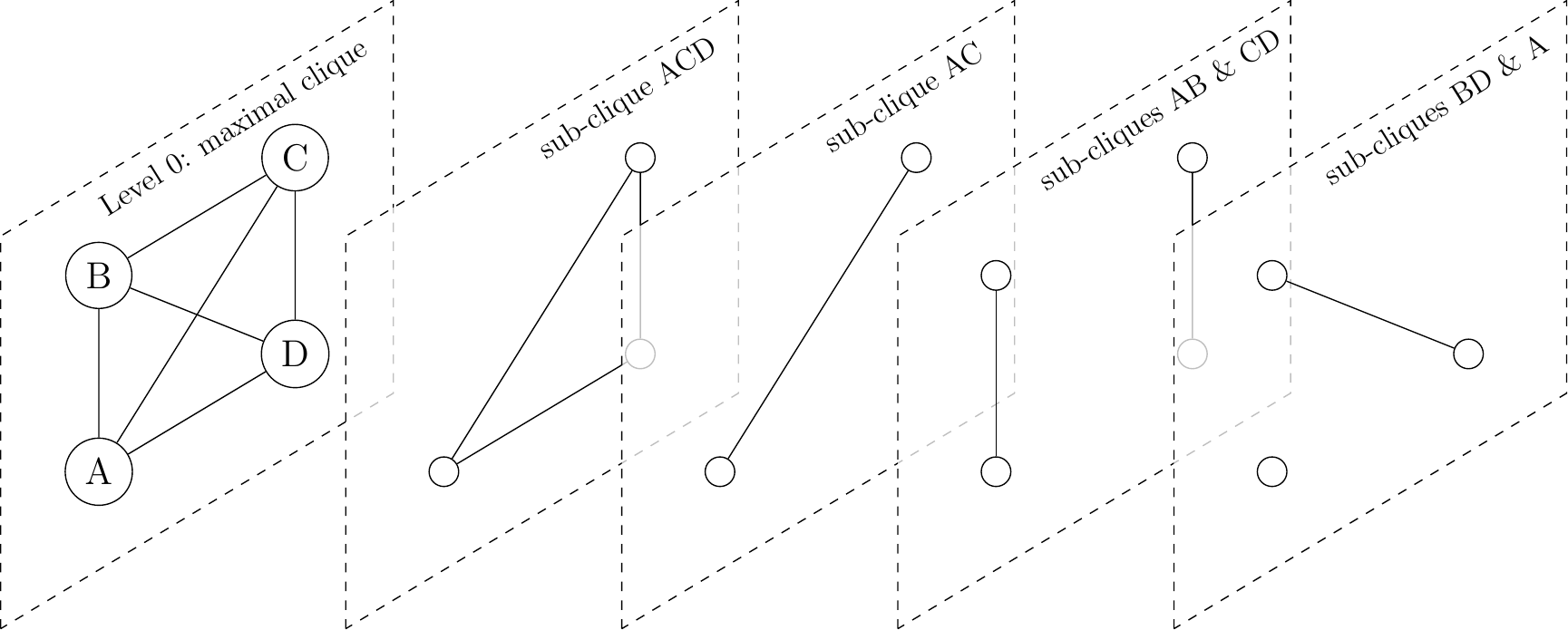}
  \caption{An example of sub-cliques as multi-edges: the maximal clique ABCD (far left) with all its sub-cliques shown in Figure \ref{fig:biadjacency-example-sub-clique}. }
  \label{fig:sub-clique-as-multi-edges}
\end{figure}

Figure \ref{fig:sub-clique-as-multi-edges} illustrates a multi-edge interpretation of the maximal clique ABCD of Figure \ref{fig:biadjacency-example-sub-clique} with all its sub-clique as multi-edges. Essentially, when disconnecting B from ABCD, the result is a maximal clique ACD, which has a sub-clique of the same size and a singleton clique B. However, when disconnecting A from ABCD, A has the sub-cliques ACD, AC, AB, and A. Hence, one must choose which of the sub-cliques to become maximal alongside BCD, either ACD, AC, AB, or disconnecting all leaving the single-node clique A. If ACD is chosen, then AC, CD and A are also sub-cliques of the former; however, AB is not. If AB is retained, \(\Z\) looses its decomposability while \(\G\) is unchanged, a direct result from the mapping in \eqref{eq:mapping-adj-biadj}.

The choice of which sub-clique to designate as maximal after a disconnection could be large. At each step, the contents and sizes of sub-cliques might differ to a large extent. Nonetheless, by their intrinsic nature, decomposable graphs favour large connected components, as in maximal cliques. To mimic this tendency while avoiding the heavy work of accounting for all combinatorially possible sub-cliques, we therefore take advantage of the continuity of the affinity parameters in choosing the sub-clique with the largest weight. We term this process "a promotion" and define it as follows.

\begin{mydef}[Promoting a sub-clique to maximal] \label{def:promoting-sub-cliques} Fallowing the settings of Definition \ref{def:clique-dependent-bipartite-graph} and Propositions \ref{prop:moved-on-Z}, let \(S_{(\theta'_{s}, \theta_{i})}\) be the set of separators contained in \(\theta'_{s} \in \C\) that contain \(\theta_{i}\), such that
  \begin{equation}\label{eq:junction-tree-edges-promot}
    S_{(\theta'_{s}, \theta_{i})} := \{\theta'_{s} \cap x: x \in \neig(\theta'_{s}, \Tr{i}{\C})\}.
  \end{equation}
  
  Promote the sub-clique \(\theta'_{o(s)}\) to maximal if \(\theta_{i}\) disconnects from \(\theta'_{s}\), where\begin{equation}\label{eq:sub-clique-choice}
  o(s):=\arg\max_{k \in \N} \{\theta'_{k} \in \neig(\theta'_{s}, \Tr{i}{\nC}): S_{(\theta'_{s}, \theta_{i})}\subset \theta'_{k},\; \deg(\theta'_{k}, T)=1 \}.
\end{equation}
\end{mydef}

In \eqref{eq:sub-clique-choice}, the result can be the empty set when no sub-cliques exist or none satisfy the conditions. The condition in \eqref{eq:sub-clique-choice} are a result of Propositions \ref{prop:moved-on-Z}.

Definition \ref{def:promoting-sub-cliques} is used in parallel with a disconnect move from a maximal clique. In the connect move of Section \ref{sec:connect-node}, a sub-clique could become maximal, however, it is a direct result of the connect move. In this case, no promotion occurs. The definition permits a promotion provided the separator set \(S_{(\theta'_s,\theta_i)}\) stays intact in a second maximal clique, and pins down the choice of which sub-clique to promote to one, if any. This streamlines the Markov update step. The next section elaborates on how the disconnect move and the promotion affect the junction graph.

\subsection{The disconnect move} \label{sec:disconnect-move}

The conditions for a node's disconnect move in \(\Z\) differs between sub-cliques and maximal cliques. No conditions are imposed on sub-cliques, as seen in \(\Tbd{i}{\nC}\), of Proposition \ref{prop:moved-on-Z}. For maximal cliques, the set \(\Tbd{i}{\C}\) is more complicated. The modification of the junction graph \(T\) is straightforward in both cases. Starting with \(T = (\Theta', \Ep)\), where \(\Theta'=\C\cup \nC\), disconnecting \(\theta_{i}\) from \(\theta'_{s}\in \Tbd{i}{\C}\cup \Tbd{i}{\nC}\) in \(\Z\) would result in the junction graph \(T'= (\C'\cup \nC', \Ep')\), with \(T'(\C')\) being the junction tree, by the following modifications:

\begin{itemize}[leftmargin=*]
\item in case \(\theta'_{s} \in \Tbd{i}{\nC}\): remove all edges \((x, \theta'_{s}) \in \Ep\) if \(\theta'_{s}\setminus \{\theta'_{i}\} = \{\emptyset\}\).
\item in case \(\theta'_{s} \in \Tbd{i}{\C}\):
  \begin{enumerate}[(a)]
  \item suppose \(\theta'_{o(s)} \in \neig(\theta'_{s}, \Tr{i}{\nC})\) is promoted to maximal (Def. \ref{def:promoting-sub-cliques}):
    \begin{enumerate}[i.]
    \item add \(\theta'_{o(s)}\) to \(\C\) and keep its edge if \(|\theta'_{o(s)}|>1\);
    \item discard sub-cliques \(\{x \in \neig(\theta'_{s}, \Tr{i}{\nC}): x \not\subseteq \theta'_{o(s)}\}\), and rewire other sub-clique edges \((x, \theta'_{s})\in \Ep\) for \(x \in \neig(\theta'_{s}, \Tr{i}{\nC})\) to \((x, \theta'_{o(s)})\);
    \item rewire all \(T_{\C}\) edges in \(S_{(\theta'_{s}, \theta_{i})}\) (see Eq. \eqref{eq:junction-tree-edges-promot}), if any, to \(\theta'_{o(s)}\).
    \end{enumerate}
    \item if \(\theta'_{s}\setminus \{\theta_{i}\}\subset x \) for \(x\in \neig(\theta'_{s}, T_{\C})\), remove \(\theta'_{s}\) from \(\C\) and rewire all its edges to \(x\).
  \end{enumerate}
\end{itemize}

The tree update moves in (a) must be done prior to (b), since the latter will rewire some edges in (a). The maximal clique disconnect tree-update moves above correspond to the single-edge disconnect move of \citet{Green01032013} when the clique is of size two, and to a subset of the multi-edge disconnect moves otherwise. 

\begin{figure}[ht]
  \centering    
  \subfloat[disconnect A from ABCD to form AB]{
        \begin{tikzpicture}[scale=0.5, transform shape, thick]
        \tikzstyle{every node}=[font=\large]
        \node[circle, draw] (a) at (0,0){A};
        \node[circle, draw] (b) at (0,2){B};
        \node[circle, draw] (c) at (2,2){C};
        \node[circle, draw] (d) at (2,0){D};

        \node[circle, draw] (e) at (4,2){E};
        \node[circle, draw] (f) at (4,0){F};

        \node[circle, draw] (g) at (6,2){G};
        \node[circle, draw] (h) at (8,0){H};
        \node[circle, draw] (i) at (8,2){I};
        
        \node (i1) at (8,2){};
        
        \draw (c) -- (b) -- (d);
        \draw (a) -- (b);
        \draw (c) --(f) -- (d) --(c);
        \draw (c) -- (e) -- (f) -- (g) -- (h) --(f);
        \draw (h) -- (i);
        
      \end{tikzpicture}\label{fig:disconnecting-single-clique-node-A}}
\subfloat[corresponding new junction graph]{
  \begin{tikzpicture}[scale=0.35, transform shape, thick]
        \tikzstyle{every node}=[font=\large]
        \node[circle, draw, red] (abcd1) at (-14,0){ABCD};
        \node[circle, draw, red] (cdf1) at (-10,0){CDF};
        \node[circle, draw, dashed] (ab1) at (-16,2){AB};

        \draw (abcd1) -- (cdf1);
        \draw[dotted] (cdf1) -- (-8,0);
        \node[circle, draw, dashed] (cf1) at (-8,-2){CF};
        \node[circle, draw, dashed] (cd1) at (-12,2){CD};
        \node[circle, draw, dashed] (a1) at (-14,2){A};
        \node[circle, draw, dashed] (d1) at (-12,-2){ACD};
        \node[circle, draw, dashed] (bd1) at (-16,-2){BD};
        \node[circle, draw, dashed] (abc1) at (-17,0){AC};
        \draw[dashed] (a1) -- (abcd1) -- (ab1);
        \draw[dashed] (bd1) -- (abcd1) -- (d1);
        \draw[dashed] (cdf1) -- (cf1);
        \draw[dashed] (abcd1) -- (cd1) -- (cdf1);
        \draw[dashed] (abc1) -- (abcd1);
        
        \draw[ -> , thick, double] (-6,0) -- (-3,0);
        \node[circle, draw, red] (abcd) at (0,0){BCD};
        \node[circle, draw, red] (cdf) at (4,0){CDF};
        \node[circle, draw, red] (ab) at (-2,2){AB};
        \draw (ab) -- (abcd) -- (cdf);
        \draw[dotted] (cdf) -- (6,0);

        \node[circle, draw, dashed] (cf) at (6,-2){CF};
        \node[circle, draw, dashed] (cd) at (2,2){CD};
        \node[circle, draw, dashed] (a) at (0,2){A};
        \node[circle, draw, dashed] (bd) at (-2,-2){BD};
        \draw[dashed] (a) -- (ab);
        \draw[dashed] (bd) -- (abcd);
        \draw[dashed] (cdf) -- (cf);
        \draw[dashed] (abcd) -- (cd) -- (cdf);
       	\end{tikzpicture}\label{fig:disconnecting-single-clique-node-B}}

      \subfloat[disconnect E from CEF to form EF]{
        \begin{tikzpicture}[scale=0.5, transform shape, thick]
        \tikzstyle{every node}=[font=\large]
        \node[circle, draw] (a) at (0,0){A};
        \node[circle, draw] (b) at (0,2){B};
        \node[circle, draw] (c) at (2,2){C};
        \node[circle, draw] (d) at (2,0){D};

        \node[circle, draw] (e) at (4,2){E};
        \node[circle, draw] (f) at (4,0){F};

        \node[circle, draw] (g) at (6,2){G};
        \node[circle, draw] (h) at (8,0){H};
        \node[circle, draw] (i) at (8,2){I};
        
        \draw (a) -- (b) -- (c) --(a) --(d) --(c) --(f) -- (d) -- (b);
        \draw (c) -- (e) ;
        \draw (f) -- (g) -- (h) --(f);
        \draw (h) -- (i);
        \node (i1) at (8,2){};
      \end{tikzpicture}\label{fig:disconnecting-single-clique-node-E}}
\subfloat[corresponding new junction graph]{
  \begin{tikzpicture}[scale=0.35, transform shape, thick]
        \tikzstyle{every node}=[font=\large]
        \node[circle, draw, red] (cdf) at (4,0){CDF};
        \node[circle, draw, red] (cef) at (8,0){CEF};
        \node[circle, draw, red] (fgh) at (12,0){FGH};
        \node[circle, draw, dashed] (ce) at (10,2){EF};
        \node[circle, draw, dashed] (cf) at (6,-2){CF};
        \node[circle, draw, dashed] (cd) at (2,2){CD};
        \node[circle, draw, dashed] (d) at (2,-2){D};
        \node[circle, draw, dashed] (gh) at (10,-2){GH};

        \draw [dotted] (2,0) -- (cdf);
        \draw [dashed] (cd) -- (cdf) -- (d);
        \draw [dashed] (cdf) -- (cf) -- (cef)-- (ce);
        \draw (fgh) -- (cef) -- (cdf);
        \draw[dashed] (fgh) -- (gh);
        \draw[dotted] (fgh) -- (12, -2);
        \draw[->, thick, double] (13.5, 0) -- (16.5,0);
        \node[circle, draw, red] (cdf1) at (4+15,0){CDF};
        \node[circle, draw, dashed, blue] (cef1) at (7.5+15,1.5){CF};
        \node[circle, draw, red] (fgh1) at (8+15,0){FGH};
        \node[circle, draw, red] (ce1) at (6+15,2){EF};
        \node[circle, draw, dashed] (cf1) at (5+15,-2){CF};
        \node[circle, draw, dashed] (cd1) at (2+15,2){CD};
        \node[circle, draw, dashed] (d1) at (2+15,-2){D};
        \node[circle, draw, dashed] (gh1) at (7+15,-2){GH};
        
        \draw [dotted] (2+15,0) -- (cdf1);
        \draw [dashed] (cd1) -- (cdf1) -- (d1);
        \draw [dashed] (cf1) -- (cdf1) -- (cef1);
        \draw (fgh1) -- (cdf1) -- (ce1);
        \draw[dashed] (fgh1) -- (gh1);
        \draw[dotted] (fgh1) -- (8+15, -2);
      \end{tikzpicture}\label{fig:disconnecting-single-clique-node-F}}

      \subfloat[disconnecting C from ABCD to form ACD]{
        \begin{tikzpicture}[scale=0.5, transform shape, thick]
        \tikzstyle{every node}=[font=\large]
        \node[circle, draw] (a) at (0,0){A};
        \node[circle, draw] (b) at (0,2){B};
        \node[circle, draw] (c) at (2,2){C};
        \node[circle, draw] (d) at (2,0){D};

        \node[circle, draw] (e) at (4,2){E};
        \node[circle, draw] (f) at (4,0){F};

        \node[circle, draw] (g) at (6,2){G};
        \node[circle, draw] (h) at (8,0){H};
        \node[circle, draw] (i) at (8,2){I};
        
        \node (i1) at (8,2){};
        
        \draw (c) -- (a) -- (d);
        \draw (d) -- (a) -- (b) --(d);
        \draw (c) --(f) -- (d) --(c);
        \draw (c) -- (e) -- (f) -- (g) -- (h) --(f);
        \draw (h) -- (i);
        
      \end{tikzpicture}\label{fig:disconnecting-multi-clique-node-A}}
\subfloat[corresponding new junction graph]{
  \begin{tikzpicture}[scale=0.35, transform shape, thick]
        \tikzstyle{every node}=[font=\large]
        \node[circle, draw, red] (abcd1) at (-14,0){ABCD};
        \node[circle, draw, red] (cdf1) at (-10,0){CDF};
        \node[circle, draw, dashed] (ab1) at (-16,2){AB};

        \draw (abcd1) -- (cdf1);
        \draw[dotted] (cdf1) -- (-8,0);
        \node[circle, draw, dashed] (cf1) at (-8,-2){CF};
        \node[circle, draw, dashed] (cd1) at (-12,2){CD};
        \node[circle, draw, dashed] (a1) at (-14,2){A};
        \node[circle, draw, dashed] (d1) at (-12,-2){ACD};
        \node[circle, draw, dashed] (bd1) at (-16,-2){BD};
        \node[circle, draw, dashed] (abc1) at (-17,0){AC};
        \draw[dashed] (a1) -- (abcd1) -- (ab1);
        \draw[dashed] (bd1) -- (abcd1) -- (d1);
        \draw[dashed] (cdf1) -- (cf1);
        \draw[dashed] (abcd1) -- (cd1) -- (cdf1);
        \draw[dashed] (abc1) -- (abcd1);
        
        \draw[ -> , thick, double] (-6,0) -- (-3,0);
        \node[circle, draw, red] (abcd) at (0,0){ABD};
        \node[circle, draw, red] (cdf) at (4,0){CDF};
        \node[circle, draw, red] (d1) at (2,2){ACD};
        \draw (cdf) -- (d1) -- (abcd);
        \draw[dotted] (cdf) -- (6,0);
        \node[circle, draw, dashed] (abc) at (0,3){AC};
        \node[circle, draw, dashed] (ab) at (-2,-2){AB};
        \node[circle, draw, dashed] (cf) at (6,-2){CF};
        \node[circle, draw, dashed] (cd) at (4.5,2.5){CD};
        \node[circle, draw, dashed] (a) at (-2,2){A};
        \node[circle, draw, dashed] (bd) at (2,-2){BD};
        \draw[dashed] (cf) -- (cdf) -- (cd) -- (d1) -- (abc);
        \draw[dashed] (a) -- (abcd) -- (ab);
        \draw[dashed] (bd) -- (abcd);
      \end{tikzpicture}\label{fig:disconnecting-multi-clique-node-B}}
    
    \caption{Examples of junction graph disconnect moves of Figure \ref{fig:biadjacency-example-sub-clique}.
      Top row: disconnecting A from ABCD and promoting AB to become maximal, BCD is still maximal, thus applying (a).\{i,ii\} of the disconnect moves. Middle row: the case when a maximal clique becomes sub-maximal, disconnecting E from CEF and promoting EF, CF is not maximal. Bottom row: disconnecting C from ABCD and promoting ACD, ABD is still maximal.}
    \label{fig:disconnecting-single-clique-node}
\end{figure}
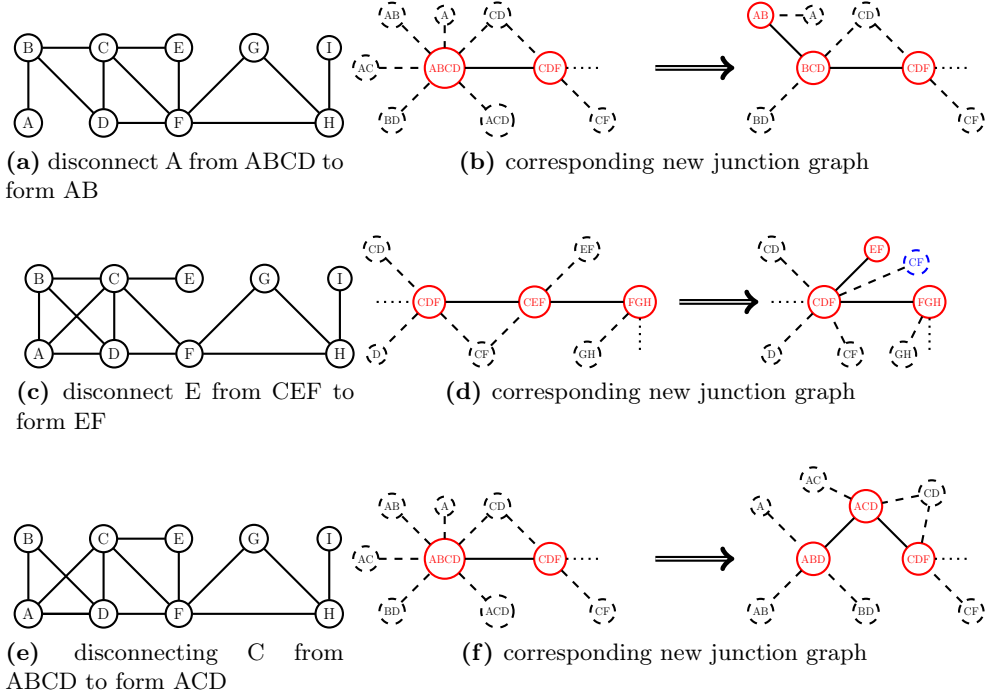

Figure \ref{fig:disconnecting-single-clique-node} is a graphical illustration of the junction graph updates for the disconnect move using the example in Figure \ref{fig:biadjacency-example-sub-clique}. The first two figure rows illustrate the disconnection of single-clique nodes where (a).iii is not needed. The last figure row is the case of disconnecting multi-clique node C from ABCD (\(\theta'_{\color{red} 1}\)), while promoting the sub-clique ACD (\(\theta'_{8}\)) to maximal.

For a complete list of possible disconnections in Figure \ref{fig:biadjacency-example-sub-clique}, and the corresponding possible promotions, refer to Table \ref{tab:list-node-disconnect-promotion} of the supplementary materials. Most disconnections do not necessarily result in new maximal cliques.

\section{The Markov chain Monte Carlo sampler} \label{sec:Markov-size-varying-steps}
One of the benefits of the bipartite representation is that it allows an easy construction of a conditional joint distribution. To show this, following the notations in Proposition \ref{prop:moved-on-Z}, and in analogy to the conditional distribution in Section \ref{sec:tree-depend-bipart}, the \(n+1\) Markov update step for \(z^{(n+1)}_{ki}\) conditional on the current configuration \(\Z^{(n)}(T)\), including that of \(z^{(n)}_{ki}\) is:
  \begin{equation}
    \P(z^{(n+1)}_{ki} = 1 \mid \Z^{(n)}(T))= 
    \begin{cases}   
 f(\theta'_k, \theta_i) & \text{if }\theta'_k \in \Tbd{i}{\C}\cup \Tbd{i}{\nC}\cup \Tnei{i}{} \\ 
z^{(n)}_{ki} & \text{ otherwise,}
\end{cases}
\end{equation}
for some integrable function \(f\), where \(\Z(T)\) is the graph \(\Z\) induced by \(T\).

The Markov nature of decomposable graphs forces nodes to establish their clique connections in \(\Z\) via a path over \(T\). For example, a node \(\theta_{i}\) initially connects to clique \(\theta'_{s_1}\); attempts unsuccessfully to connect to neighbouring cliques-nodes of \(\theta'_{s_1}\) in \(T\); with a successful connection to \(\theta'_{s_2}\); \(\theta_{i}\) attempts the neighbours of \(\theta'_{s_2}\) that are not yet attempted, and so on. This results in \(\Tr{i}{}\), which defines the successful connection path of \(\theta_{i}\), the unsuccessful attempts are defined by \(\Tnei{i}{}\). This path construction is achieved through many Markov iterations on the junction graph \(T\). Nonetheless, omitting the iteration notation, a conditional joint distribution of the connection path of \(\theta_{i}\) in \(T\) is
\begin{equation}\label{eq:joint-distribution-general}
    \P(\z_{.i} \mid \Z_{-(.i)}(T))  = \Bigg \{ \prod_{x \in \ver(\Tr{i}{})} f(x, \theta_{i}) \Bigg \}\Bigg \{ \prod_{x\in \Tnei{i}{}} 1-  f(x, \theta_{i})\Bigg \},
\end{equation}
where and \(\Z_{-(.i)}\) is \(\Z\) excluding the \(i\)-th column. For \(\deltanei_{ki} = 1\) if \(\theta'_k  \in \Tnei{i}{}\), otherwise 0, it simplifies to 
\begin{equation}
  \P(\z_{.i} \mid \Z_{-(.i)}(T)) = \prod_{k=1}^{\mid \Theta'\mid }\Big \{ f(\theta'_{k}, \theta_{i}) \Big \}^{z_{ki}}\Big \{ 1-  f(\theta'_{k}, \theta_{i})\Big \}^{(1-z_{ki})\deltanei_{ki}}.
\end{equation}

A node-driven joint distribution can also be constructed using the decomposable graph \(\G(T_{\G})\) directly, with its junction tree \(T_{\G}\). Albeit, the \(\Z\) representation is more intuitive.

Regarding the proposal distribution, and as a consequence of \eqref{eq:joint-distribution-general}, for a given \(\theta_{i}\), some clique-nodes in \(\Tbd{i}{\C}\cup \Tbd{i}{\nC}\cup \Tnei{i}{}\) can be updated independently and simultaneously. For example, disconnecting from \(\Tbd{i}{\nC}\) and connecting to \(x \in \Tnei{i}{\nC}\) where \(\theta_{i}\in \pa(x)\), can all be made simultaneously. In such case, the proposal is uniform proportional to \(1/|\Theta|\), for each.

For some clique-nodes in \(\Tnei{i}{}\) that are neighbouring ones in \(\Tbd{i}{\C}\), one can either connect to the former or disconnect from the latter, not both simultaneously. For such cases, let \(M\) be the set of clique-nodes in \(\Tbd{i}{\C}\) that are neighbouring in \(T\) to a clique-node in \(\Tnei{i}{\C}\cup \{ x \in \Tnei{i}{\nC}: \theta_{i} \notin \pa(x)\}\). Select \(m\) uniformly from \(M\), and disconnect them with proposal probability \(m!(|M|- m)!/(|\Theta||M|!)\). Moreover, connect to all clique-nodes in \(\Tnei{i}{\C}\cup \{ x \in \Tnei{i}{\nC}: \theta_{i} \notin \pa(x)\}\), that are not neighbouring to the \(m\) selected ones, with the same proposal. For all other clique-nodes the proposal is uniform proportional to \(1/|\Theta|\).

\section{Discussion}
The sub-clustering interpretation in decomposable graphs allows for novel applications in behaviour type of data. For example, breaking up cliques into their sub-clusters can model certain dynamics in economics, such as mergers and acquisitions, where firms purchase units within others. Such interpretation can also be applied in biology, sports and other fields. With regards to graphical models, a factorization theorem in terms of \(\Z\) rather than \(\G\) is possible, see \cite[Sec 4.2]{Elmasri2017a}. In such case, the factorization is only influenced by the maximal cliques and not their sub-clusters. Nonetheless, the node-driven update scheme with sub-clusters might improve the convergence of the model, where more flexible updates are possible.

The flexibility and depth that are gained by accounting for sub-cliques comes with extra complexities, primarily related to the dynamics between cliques and sub-cliques. It is not clear how these dynamics should be structured; for example, when disconnecting a node from a clique, does it also disconnect from all its sub-cliques? It does not in this work, unless through a promotion move. Nonetheless, other schemes are possible, for example a penalty scheme.

The clustering mechanism proposed in this work does not depend on choosing the correct number of clusters, nor on choosing a proper clustering distance. It adopts a fixed size bipartite graph \(\Z\); hence, as long as the number of clique-nodes is larger than the number of nodes, one can potentially infer the correct number of maximal cliques. All other clique-nodes are labelled as sub-clusters. A possible improvement is a method for choosing the number of desired sub-clusters. For example, adopting a sub-clustering framework that is in between the proposed interpretation of Section \ref{sec:sub-cliques-as-multi-edges}, and the initial representation in Section \ref{sec:tree-depend-bipart}. Clique-nodes are initially treated as latent communities representing maximal cliques with a single Markov update scheme (\ref{eq:Update-step-simple}). This also amounts to a notion of sub-clustering for non-maximal cliques, with a less complex update steps, though with a different interpretation. Here, sub-maximal cliques are potentially maximal as more nodes are added to the model, and thus are only temporary sub-clusters.

\bibliographystyle{chicago}  
\bibliography{references}

\begin{appendices}
\addtocontents{toc}{\protect\setcounter{tocdepth}{0}}
\section{Graph perturbations independent of junction trees}
In Proposition \ref{prop:moved-on-Z}, the boundary and neighbouring sets can be specified without conditioning on a junction tree, as follows

\begin{equation}
  \begin{aligned}
  \label{eq:nei-bound-sets-ind-of-tree}
  \Tbd{i}{\C} &= \Big \{\theta'_s \in \C: \theta_{i} \in \theta'_{s}, S_{(\theta'_{s}, \theta_{i})} \subseteq \theta'_{k}, \theta'_{k} \in \Theta' \Big\},\\
    \Tbd{i}{\nC} &= \Big \{\theta'_s \in \nC: \theta_{i} \in \theta'_{s} \Big\},\\
    \Tnei{i}{\C} &= \Big \{\theta'_s \in \Theta' : \theta_{i} \not\in \theta'_{s}, \theta'_{s}\cap \{ \C \setminus \theta'_{s}\} \subseteq \theta'_{k}\text{ for } \theta'_{k}\in \Tr{i}{\C} \Big\},\\
    \Tnei{i}{\nC} &= \Big \{\theta'_s \in \Theta' : \theta_{i} \not\in \theta'_{s}, \theta'_{s}\cap \C \subseteq \theta'_{k}\text{ for } \theta'_{k}\in \Tr{i}{\C} \Big\}.\\
  \end{aligned}
\end{equation}

Essentially, all the conditions above build on the running intersection property of POS's, where \(\S_{(\theta'_{s}, \theta_{i})}\) is the set of separators in \(\theta'_{s}\) that contain \(\theta_{i}\).
\section{List of possible disconnections of Figure \ref{fig:biadjacency-example-sub-clique}}

\begin{table}[ht!]
  \centering
  \caption{List of possible disconnect moves and sub-cliques promotion of example in Figure \ref{fig:biadjacency-example-sub-clique}}
  \label{tab:list-node-disconnect-promotion}
  \begin{tabular}[ht!]{|l |l |l |l |}
    \hline 
    \(\theta_{i}\) & \(\theta'_{s}\)& \(S_{(\theta'_{s}, \theta_{i})}\) & \(\{x \in \nC: S_{(\theta'_{s}, \theta_{i})} \subset x, \deg(x, T)=1\} \) \\
    \hline
    A & ABCD & \{\(\emptyset\)\} & \{A, AB, AC, ACD\} \\
    B & ABCD & \{\(\emptyset\)\} & \{AB, BD\} \\
    E & CEF & \{\(\emptyset\)\} & \{EF\}\\
    G & FGH & \{\(\emptyset\)\} & \{GH\} \\
    I & HI & \{\(\emptyset\)\} & \{HI\} \\
  C & ABCD & \{CD\}     & \{ACD\}           \\      
  C & CDF  & \{CD, CF\} & \{\(\emptyset\)\} \\
  C & CEF  & \{CF\}     & \{\(\emptyset\)\}            \\
  D & ABCD & \{CD\}     & \{ACD\}           \\
  D & CDF  & \{CD\}     & \{\(\emptyset\)\} \\
  F & CDF  & \{CF\}     & \{\(\emptyset\)\} \\
  F & CEF  & \{CF,F\}   & \{\(\emptyset\)\} \\
  F & FGH  & \{F\}      & \{\(\emptyset\)\} \\
  H & FGH  & \{H\}      & \{GH\}            \\
  H & HI   & \{H\}      & \{HI\}            \\
  \hline
\end{tabular}
\end{table}

\end{appendices}

\end{document}

%% file: mypreamble.tex
\usepackage[english]{babel}
\usepackage{amsmath,amssymb,amsthm}
\usepackage[latin9]{inputenc}
\usepackage{tablefootnote}
\usepackage[T1]{fontenc}
\usepackage{graphicx, xcolor,tikz,arrayjobx,trimspaces,subfig}
\usepackage[labelfont=bf]{caption}
\usepackage{enumitem}           
\usepackage{enumerate,multirow}
\usepackage{soul}
\usepackage{mathtools}

\usepackage[mathlines]{lineno}             
\usepackage[round, sort, numbers, authoryear]{natbib}   
\usepackage[colorlinks=true,citecolor=blue,urlcolor=blue]{hyperref}                    

\usepackage{color}
\usepackage{marvosym}  

\usepackage{pgfplots}

\pgfplotsset{compat=1.11}
\usetikzlibrary{calc}
\usetikzlibrary{positioning}
\usetikzlibrary{decorations.pathreplacing}

\usepackage[toc, page]{appendix}

\usepackage{todonotes}   

\newtheorem{theo}{Theorem}[section]
\newtheorem{mydef}[theo]{Definition}

\newtheorem{myprop}[theo]{Proposition}

\theoremstyle{definition}

\theoremstyle{remark}
\newtheorem*{remark}{Remark} 
\numberwithin{equation}{section}

\def\P{{P}}

\def\N{\mathbb{N}}

\def\Rp{\mathbb{R}_+}

\def\Z{\mathbf{Z}}
\def\z{\mathbf{z}}

\def\treebi{\widetilde \Z}

\def\G{\mathcal{G}}
\def\C{\mathcal{C}}                 
\def\nC{\overline{ \mathcal{C}}}           
\def\S{\mathcal{S}}

\def\Ep{\mathcal{E}}

\def\ver{\boldsymbol v}

\def\pa{{\textsf{pa}}}
\newcommand{\Tr}[3][]{T^{#1\mid #2}_{#3}}
\newcommand{\Tbd}[3][]{\prescript{#3}{}{T}_{\textsf{bd}}^{#1 \mid #2}}
\newcommand{\Tnei}[3][]{\prescript{#3}{}{T}_{\textsf{nei}}^{#1 \mid #2}}

\def\deg{{\sf deg}}
\def\neig{{\sf nei}}

\def\deltanei{{\delta^{\textsf{nei}}}}

\def\textcite{\citet}